# Convenient Analytical Solution for Vibrational Distribution Function of Molecules Colliding with a Wall

Wei Yang[1, 2, 3], Alexander V. Khrabrov[2], Igor D. Kaganovich[2] and You-Nian Wang[3]

[1]*College of Science, Donghua University, Shanghai 201620, China*

[2]*Plasma Physics Laboratory, Princeton University, Princeton, New Jersey 08543, USA*

[3]*School of Physics, Dalian University of Technology, Dalian 116024, China*

**Abstract**

We study formation of the Vibrational Distribution Function (VDF) in a molecular gas at low pressure, when vibrational levels are excited by electron impact and deactivated in collisions with walls and show that this problem has a convenient analytical solution that can be used to obtain VDF and its dependence on external parameters. The VDF is determined by excitation of vibrational levels by an external source and deactivation in collisions with the wall. Deactivation in wall collisions is little known process. However, we found that the VDF is weakly dependent on the functional form of the actual form of probability $\gamma_{\upsilon'\to\upsilon}$ for a vibrational number $\upsilon'$ to transfer into a lower level $\upsilon$ at the wall. Because for a given excitation source of vibrational states, the problem is linear the solution for VDF involves solving linear matrix equation. The matrix equation can be easily solved if we approximate probability, in the form: $\gamma_{\upsilon'\to\upsilon} = (1/\upsilon')\theta(\upsilon'-\upsilon)$. In this case, the steady-state solution for VDF( $\upsilon$ ) simply involves a sum of source rates for levels above $\upsilon$, with a factor of $1/(\upsilon+1)$. As an example of application, we study the vibrational kinetics in a hydrogen gas and verify the analytical solution by comparing with a full model.

Vibrational kinetics is important for many applications such as negative ion sources, plasma processing, and in the general area of surface (heterogeneous) catalysis. In this letter we are focused on the pressure range, generally, below 100 mTorr where the Vibrational Distribution Function (VDF) of a molecular gas is determined by balance of electron-impact excitation and deactivation in collisions with the chamber wall. As an example, we study the vibrational kinetics of $H_2$ with application to negative ion sources, but the treatment can be employed in other applications.

Negative Hydrogen Ion Sources (NHIS) are needed for producing fast neutral beams for heating plasma in nuclear fusion reactors. Development of NHIS for such neutral-beam heating systems requires an efficient generation of $H^-$ ions in low-pressure plasmas. A detailed review of negative hydrogen ion production mechanisms can be found in a recent paper by Bacal *et al*. [1]. The presence of vibrationally excited $H_2$ molecules in hydrogen discharges enhances the production of negative ions through the process of dissociative attachment [2,3]. A population of $H_2$ molecules in higher vibrational states is necessary to increase production of negative ions [4,5]. Therefore, since 1970s, the studies were focused on the vibrational kinetics of $H_2$ plasmas [6,7]. The production and loss mechanisms for excited states include volume reactions and interactions with chamber walls. An important process is deactivation of vibrational level at the wall surface to form a lower state, i.e., relaxation $\upsilon' \rightarrow \upsilon$, with $0 \le \upsilon < \upsilon'$. The probabilities $\gamma_{\upsilon' \rightarrow \upsilon}$ for this process were originally calculated by Hiskes and Karo [8] using molecular-trajectory simulations. Experimental measurements of repopulation probability distributions were initially reported by Stutzin *et al.* in unpublished conference proceedings [9]. Their data was published by Hiskes and Karo [10] and has been since widely adopted in modeling $H_2$ discharges. Examples of VDF simulations using global-model approach are given in publications [11-15]. In such calculations, it is difficult to decouple the combined effects of different kinetic processes responsible for forming the VDF. Hence it is not transparent which factors predominantly affect the shape of the resulting $H_2$ vibrational spectrum.

In this letter, we develop a convenient analytical solution for vibrational distribution function of molecules colliding with a wall; we call it a Reduced Linear Model (RLM) of vibrational kinetics. RLM allows for a fast and straightforward VDF calculation and also enables analysis of contribution of different kinetic processes into the VDF formation.

In our previous study [16], a benchmarked and validated Global Model for Negative Hydrogen Ion Source (GMNHIS) has been developed. We benchmarked the GMNHIS against another independently developed code, Global Enhanced Vibrational Kinetic Model (GEVKM) [14] and validated the GMNHIS using experimental measurement data obtained in an electron cyclotron resonance (ECR) discharge [17]. The GMNHIS code implements a quite comprehensive reaction set for vibrational kinetics in $H_2$. For the present work, all reaction rates related to the creation and loss of vibrational levels are evaluated according to GMNHIS, and reactions with small contributions are neglected. The RLM of vibrational kinetics is derived for low-pressure NHIS through (a) reducing the reaction set of vibrational kinetics of $H_2$ molecules and (b) using approximate function for the repopulation probability $\gamma_{\upsilon' \rightarrow \upsilon} = (1/\upsilon')\theta(\upsilon' - \upsilon)$. We show that such an approximation does not significantly affect the calculated VDF, but allows for significant simplifications. The simplified reaction set of vibrational kinetics of $H_2$ molecules includes electron excitation of $H_2(\upsilon = 0)$ ground-state molecules to levels $H_2(\upsilon = 1-14)$ through a resonant mechanism (called eV process) and through an indirect mechanism followed by radiative decay (called EV process), and also wall relaxation (WR process) of a vibrational level $\upsilon'$ to a lower level $\upsilon$ $(< \upsilon')$.

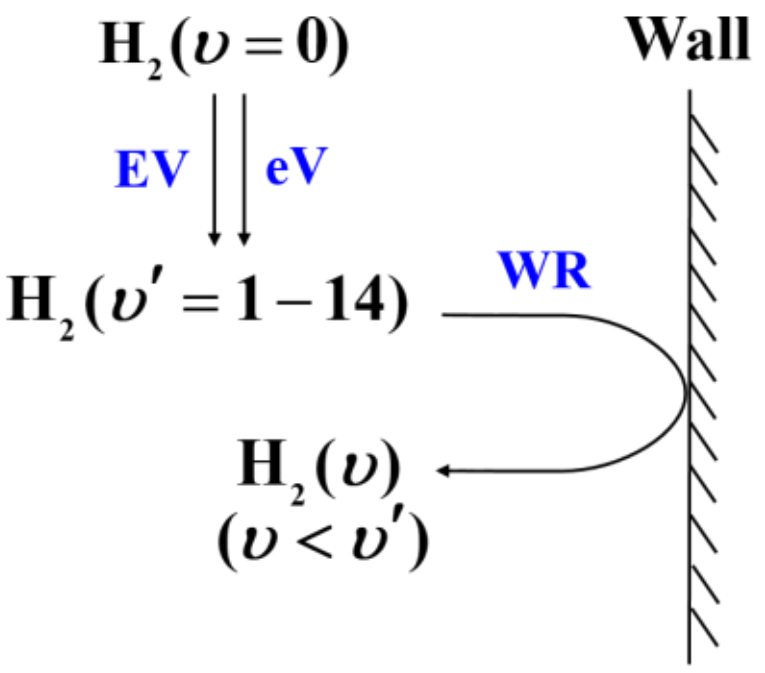

Figure 1. Schematics of the processes included in the RLM for the formation of VDF.

Figure 1 illustrates the reduced set of processes for vibrational kinetics mentioned above. The number density of molecules in a given vibrational state is linearly dependent on the ground state $H_2(\upsilon=0)$ through the source rate (eV and EV processes), and on the densities of molecules at higher vibrational levels due to wall collisions (WR process). The highest vibrational level $H_2(\upsilon=14)$ is not produced in wall collisions (because there is no higher vibrational level) and, therefore, the densities of molecules at $H_2(\upsilon=14)$ is only proportional to the density of $H_2(\upsilon=0)$. The density of the second highest vibrational level, $H_2(\upsilon=13)$, is given by a linear combination of the densities of $H_2(\upsilon=0)$ and $H_2(\upsilon=14)$, and so on. The densities of other remaining of vibrational levels can be calculated in the same way. Therefore, the densities of $H_2(\upsilon=1-14)$ satisfy a set of $N=14$ linear equations with a triangular matrix in the form

$$S_\upsilon = (k_{\mathrm{WR},\upsilon} + k_{\mathrm{out}}) n_{\mathrm{H}_2(\upsilon)} - \sum_{\upsilon'=\upsilon+1}^{N} k_{\mathrm{WR},\upsilon'} \gamma_{\upsilon'\to\upsilon} n_{\mathrm{H}_2(\upsilon')}, \tag{1}$$

where the source rate $S_\upsilon = (k_{\mathrm{eV},0\to\upsilon} + k_{\mathrm{EV},0\to\upsilon}) n_{\mathrm{H}_2(0)} n_e$ is given by the direct-excitation rate constants. In the RLM, the electrons are assumed to be at equilibrium, therefore the rate constants ($k_{\mathrm{eV},0\to\upsilon}$ and $k_{\mathrm{EV},0\to\upsilon}$) are obtained under assuming a Maxwellian electron energy distribution. The rate constant for the wall relaxation processes is given, according to Chantry [18], as

$$k_{\mathrm{WR},\upsilon} = \left[ \frac{\Lambda^2}{D} + \frac{2V}{Av_{th}} \left( \frac{1+R_\upsilon}{1-R_\upsilon} \right) \right]^{-1}. \tag{2}$$

Here, $\Lambda$ is the diffusion scale of the reactor vessel of volume *V* and surface area *A*, with gas diffusion coefficient *D*, and the mean thermal velocity $v_{th}$, both of the latter independent on the vibrational number $\upsilon$; $R_\upsilon$ is the reflection coefficient (the reflected flux is Maxwellian with a cosine-law angular dependence). The values of $R_\upsilon$ for $H_2(\upsilon = 1-14)$, in the form $R_\upsilon = \exp(-1/b_\upsilon)$, were calculated by Hiskes and Caro [8,10] (the quantities $b_\upsilon$ are collision frequencies if time is measured by the number of successive collisions with the wall). Gas-surface interactions are generally dependent on the material and temperature of the chamber wall. The cited-above molecular-trajectory calculations were performed for iron wall at 500 K and thus should be applicable to stainless-steel vacuum vessels. Note that the temperature dependence isn't strong, because the energy of vibrational quantum (on the order of 0.1 eV) is high compared with the thermal energy of wall atoms (not much higher than 0.05 eV). The repopulation probabilities are normalized as $\sum_{\upsilon''=0}^{\upsilon-1} \gamma_{\upsilon\to\upsilon''} = 1$.

The rate constant $k_{\text{out}}$ defines pumping (outflow) loss of the excited molecules out of the volume accounted for by the global model, e.g. they may escape with the outlet flow from the chamber. It must be mentioned that Eq. (2), initially derived for an enclosed chamber without any pumping, is only valid under the condition $k_{\text{out}} \ll k_{\text{WR}}$; however, such global models are often used even when the outflow is considerable such that $k_{\text{out}} \sim k_{\text{WR}}$.

Eq. (1) was introduced in Ref. [10] to study the dependence of predicted vibrational spectrum on the total repopulation probability $1 - R_\upsilon$. Numerical solution of the matrix equation (1) is trivial, starting with $\upsilon = N$ and solving for lower levels one by one. A compact analytical solution is possible if the form of $\gamma_{\upsilon'\to\upsilon}$ is simplified. Having such solution would facilitate physical insight, as well as verification of complex numerical models, and is the focus of the present work. As noted above, the measured values of $H_2$ repopulation probabilities $\gamma_{\upsilon'\to\upsilon}$ currently in common use originate from results [9] reported originally in conference proceedings. These data are used even though there is no published record of experimental conditions and error bounds for the measurements. Given that considerable uncertainties are present in other basic inputs into the model (notably the cross-section values) utilized in Eq. (1), and the fact that the zero-dimension global model is approximate itself, we show that reasonable approximation for $\gamma_{\upsilon'\to\upsilon}$ does not degrade overall accuracy of the predictions for VDF.

The experimental plots of $\gamma_{\upsilon'\to\upsilon}$ are shown in Fig. 2(a).

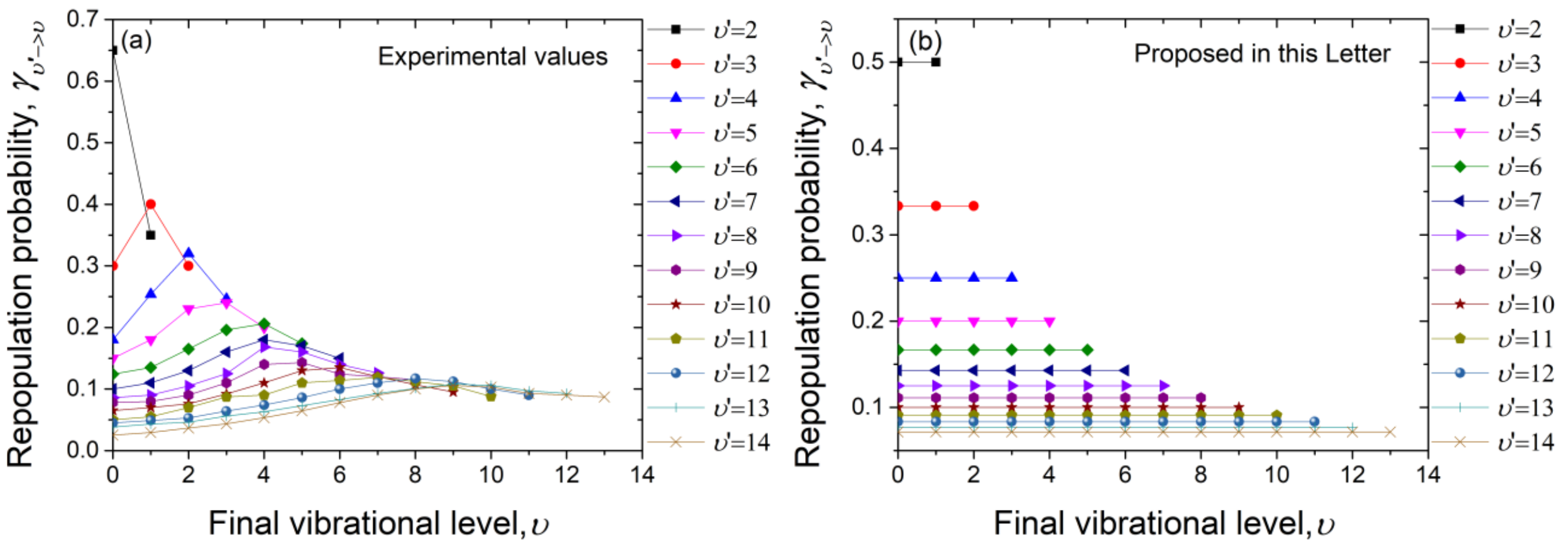

Figure 2: Repopulation probability $\gamma_{\upsilon'\to\upsilon}$ in wall collisions: (a) data provided in Ref. [9,10]; (b) approximate $\gamma_{\upsilon'\to\upsilon} = (1/\upsilon')\theta(\upsilon'-\upsilon)$ proposed in this Letter.

Because the $\gamma_{\upsilon'\to\upsilon}$ functions are smooth in the sense that the scale of the distribution is determined by the initial vibrational number $\upsilon'$, and these functions enter Eq. (1) via summation, we propose to simply replace $\gamma_{\upsilon'\to\upsilon}$ with respective average values keeping the normalization, that is:

$$\gamma_{\upsilon'\to\upsilon} = (1/\upsilon')\theta(\upsilon'-\upsilon)\,.$$

Subsuming above into Eq. (1) gives

$$S_\upsilon = (k_{\mathrm{WR},\upsilon} + k_{\mathrm{out}})n_{\mathrm{H}_2(\upsilon)} - \sum_{\upsilon'=\upsilon+1}^{N} \frac{1}{\upsilon'} k_{\mathrm{WR},\upsilon'} n_{\mathrm{H}_2(\upsilon')}. \tag{3}$$

We solve Eq. (3) analytically in Appendix A and the solution is

$$k_{\mathrm{WR},\upsilon} n_{\mathrm{H}_2(\upsilon)} = \beta_\upsilon S_\upsilon + \frac{\beta_\upsilon}{\upsilon+1}\left\{\beta_{\upsilon+1}S_{\upsilon+1} + \sum_{k=\upsilon+2}^{N} \beta_k S_k \prod_{n=\upsilon+1}^{k-1} \alpha_n\right\}, \tag{4}$$

where $\beta_\upsilon = \dfrac{k_{\mathrm{WR},\upsilon}}{k_{\mathrm{WR},\upsilon} + k_{\mathrm{out}}}$ and $\alpha_n = (\beta_n + n)/(n+1)$. In the case of pumping loss is not taken into account $\alpha_\upsilon = \beta_\upsilon = 1$, and the solution is

$$k_{\mathrm{WR},\upsilon} n_{\mathrm{H}_2(\upsilon)} = S_\upsilon + \frac{1}{\upsilon+1}\sum_{k=\upsilon+1}^{N} S_k\,. \tag{5}$$

The left-hand side of Eqs. (4) and (5) describes the volumetric loss (in $\mathrm{m}^{-3}\mathrm{s}^{-1}$) of excited molecules at the wall surface and the right hand side describes the volumetric source.

Henceforth, we will be considering the VDF predicted by Eq. (5) rather than the general solution given by Eq. (4) and we compare this solution to the results of our global-model simulation [16] with $k_{\text{out}} = 0$. As evident from the solution given by Eq. (5), the resulting effect of wall collisions is given by a response function with respect to the source, $\frac{\theta(k-\upsilon)}{\upsilon+1}$, which is a simple step function.

Figure 2(b) shows the adopted probability $\gamma_{\upsilon'\to\upsilon} = (1/\upsilon')\theta(\upsilon'-\upsilon)$ distribution for a vibrational level $\upsilon'$ to relax to lower states.

Figures 3(a) and 3(b) show the corresponding linear response obtained by solving the respective elements of the inverse matrix, $\boldsymbol{B}^{-1}$, where the elements of $\boldsymbol{B}$ are $B_{\upsilon,\upsilon'} = \delta_{\upsilon,\upsilon'} - \gamma_{\upsilon'\to\upsilon}$, corresponding to Eq. (1) with $k_{\text{out}} = 0$, for $\gamma_{\upsilon'\to\upsilon}$ data taken from the experimental study of Ref. [9] as reproduced in Ref. [10] shown in Fig. 2(a) and approximated by our model shown in Fig. 2(b), respectively. The values for $\boldsymbol{B}^{-1}$ are almost constant values for a fixed row ($\upsilon$) of $\boldsymbol{B}^{-1}$ and well approximated by $\frac{\theta(k-\upsilon)}{\upsilon+1}$. Therefore, instead of using the tabulated values of $\gamma_{\upsilon'\to\upsilon}$ from Refs. [9,10] we could conveniently replace them simply with $\gamma_{\upsilon'\to\upsilon} = (1/\upsilon')\theta(\upsilon'-\upsilon)$.

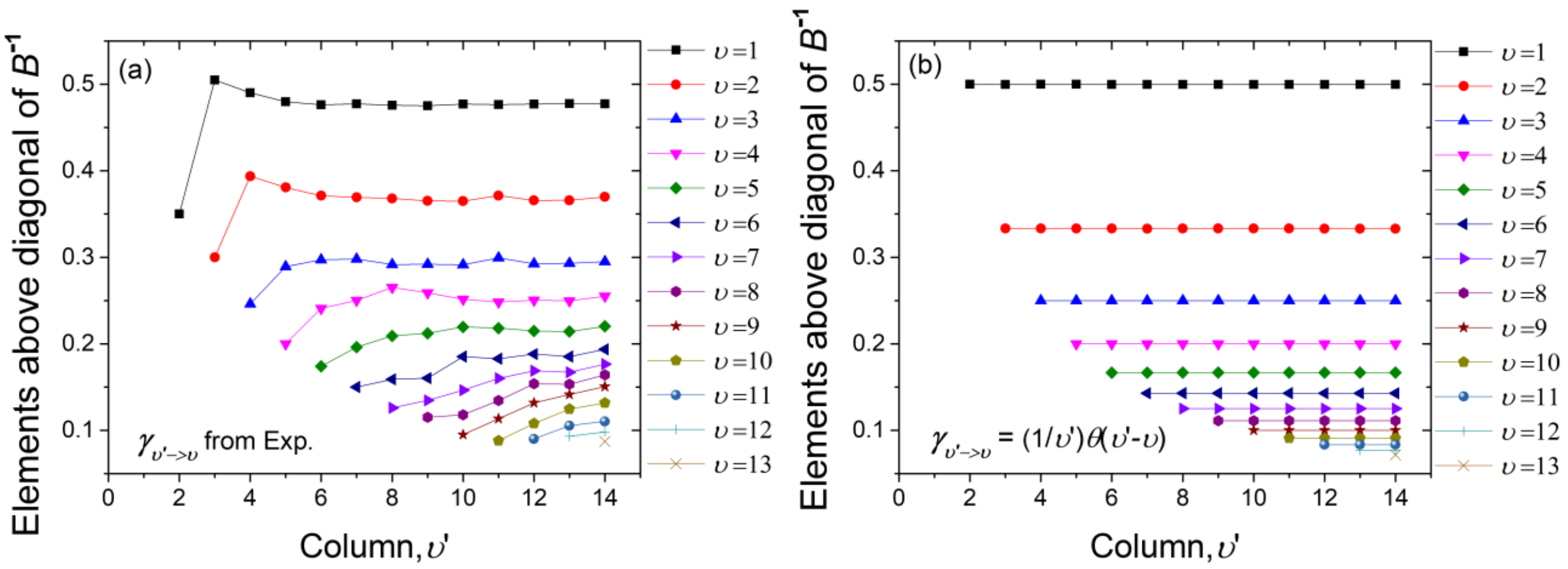

Figure 3: Values of the row elements of the corresponding inverse matrix for Eq.(1) for initial $\gamma_{\upsilon'\to\upsilon}$ (a) and approximate $\gamma_{\upsilon'\to\upsilon} = (1/\upsilon')\theta(\upsilon'-\upsilon)$ (b). The diagonal elements, equal to unity, are not shown.

In order to test the proposed analytical solution given by Eq. (5), VDFs were simulated using GMNHIS [16] with $\gamma_{\upsilon'\to\upsilon}$ from [9,10] and also $\gamma_{\upsilon'\to\upsilon} = (1/\upsilon')\theta(\upsilon'-\upsilon)$ for two values of the gas pressure:

4 mTorr and 18 mTorr (accounting only for the reduced set of reactions considered in this Letter). The simulated VDFs are shown in Fig. 4. The results clearly show that the obtained VDFs nearly indistinguishable and the proposed substitution yield rather accurate VDF.

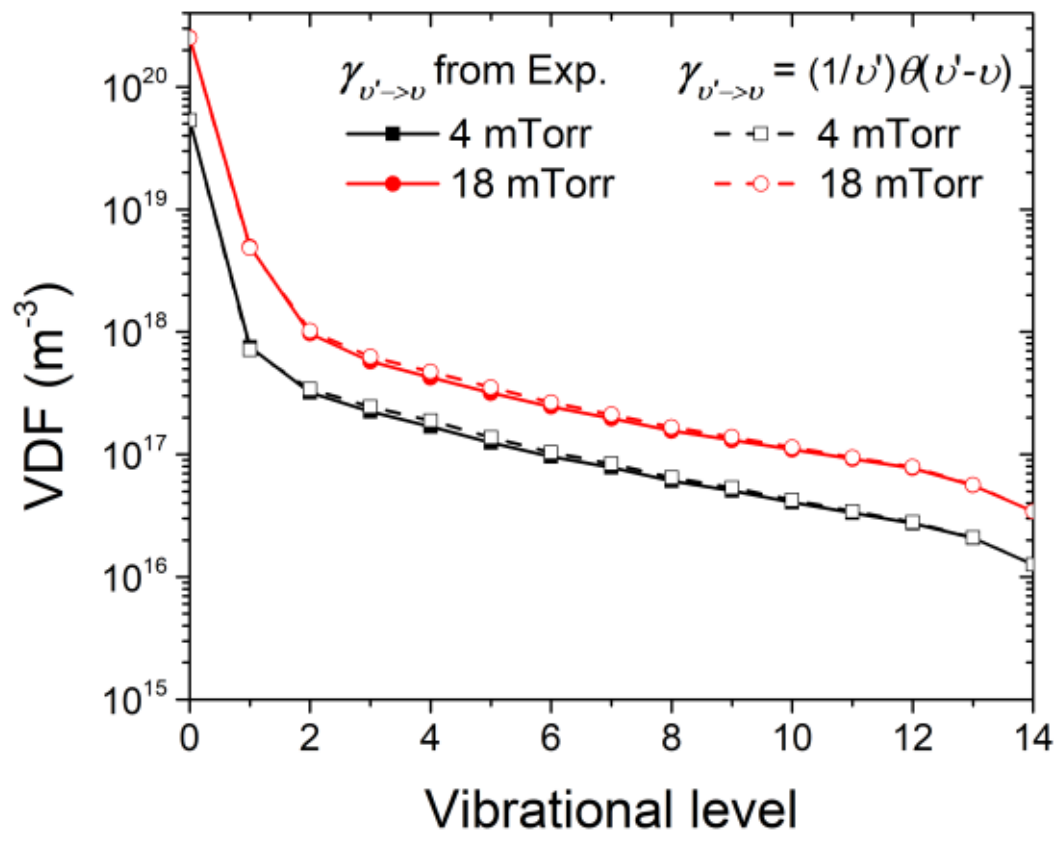

Figure 4. VDFs obtained using GMNHIS with the reduced set of reactions considered here and $\gamma_{\upsilon'\to\upsilon}$ from Refs. [9,10], compared with VDFs obtained in simulations with $\gamma_{\upsilon'\to\upsilon} = (1/\upsilon')\theta(\upsilon'-\upsilon)$ for two values of gas pressure: 4 mTorr and 18 mTorr.

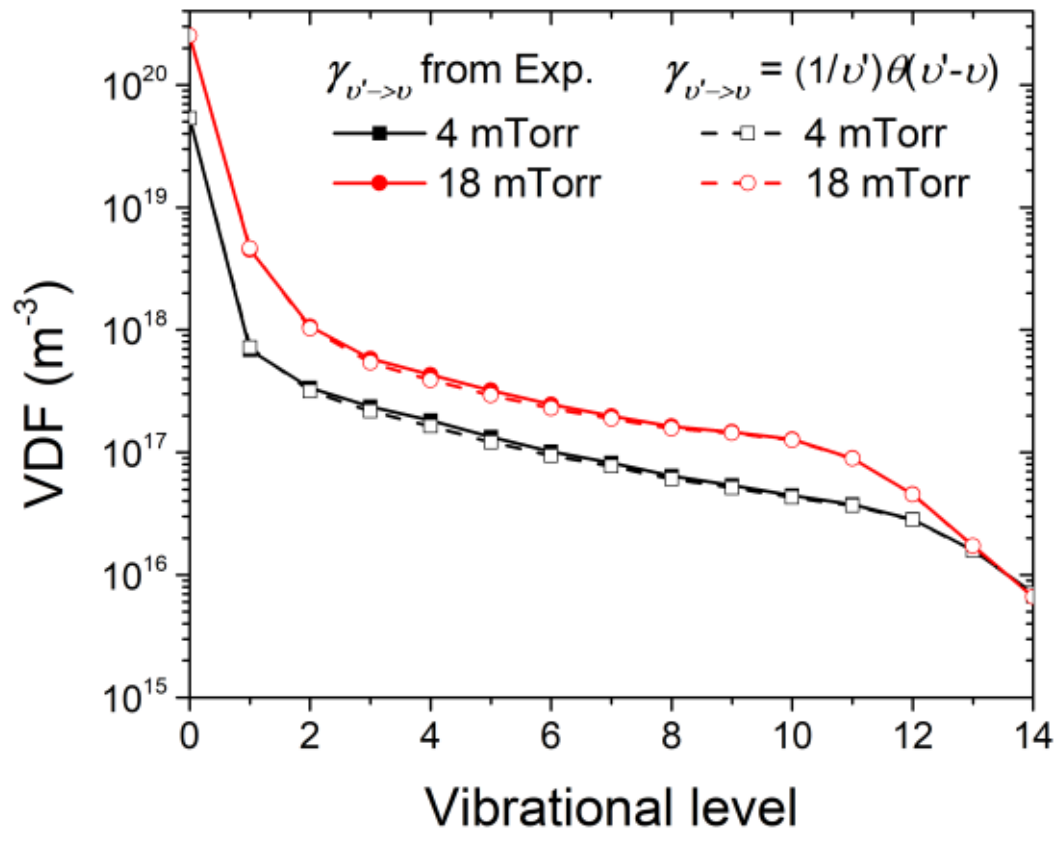

Figure 5. VDFs obtained using full GMNHIS and $\gamma_{\upsilon'\to\upsilon}$ from Refs. [9,10] and compared obtained VDF with simulations using $\gamma_{\upsilon'\to\upsilon} = (1/\upsilon')\theta(\upsilon'-\upsilon)$ for two gas pressure values: 4 mTorr and 18 mTorr.

Figure 5 also shows the VDFs obtained using full set of reactions accounted in GMNHIS with $\gamma_{\upsilon'\to\upsilon}$ from [9,10] and also $\gamma_{\upsilon'\to\upsilon} = (1/\upsilon')\theta(\upsilon'-\upsilon)$ for 4 mTorr and 18 mTorr. Again, the approximate distribution $\gamma_{\upsilon'\to\upsilon} = (1/\upsilon')\theta(\upsilon' - \upsilon)$ results in a nearly indistinguishable VDF. Comparing full simulation shown in Fig. 5 with reduced set shown in Fig. 4, we note that they agree well with each other except for the highest vibrational levels. This is because vibrational-translational relaxation in collisions with molecular hydrogen (VT) was ignored in the reduced set of equations. The VT process can affect the densities of very high vibrational states, but its effect is weakened with decreasing pressure. To demonstrate the difference Fig. 6 shows comparison of the VDFs obtained with full GMNHIS and the RLM (reduced set without taking the VT process into account).

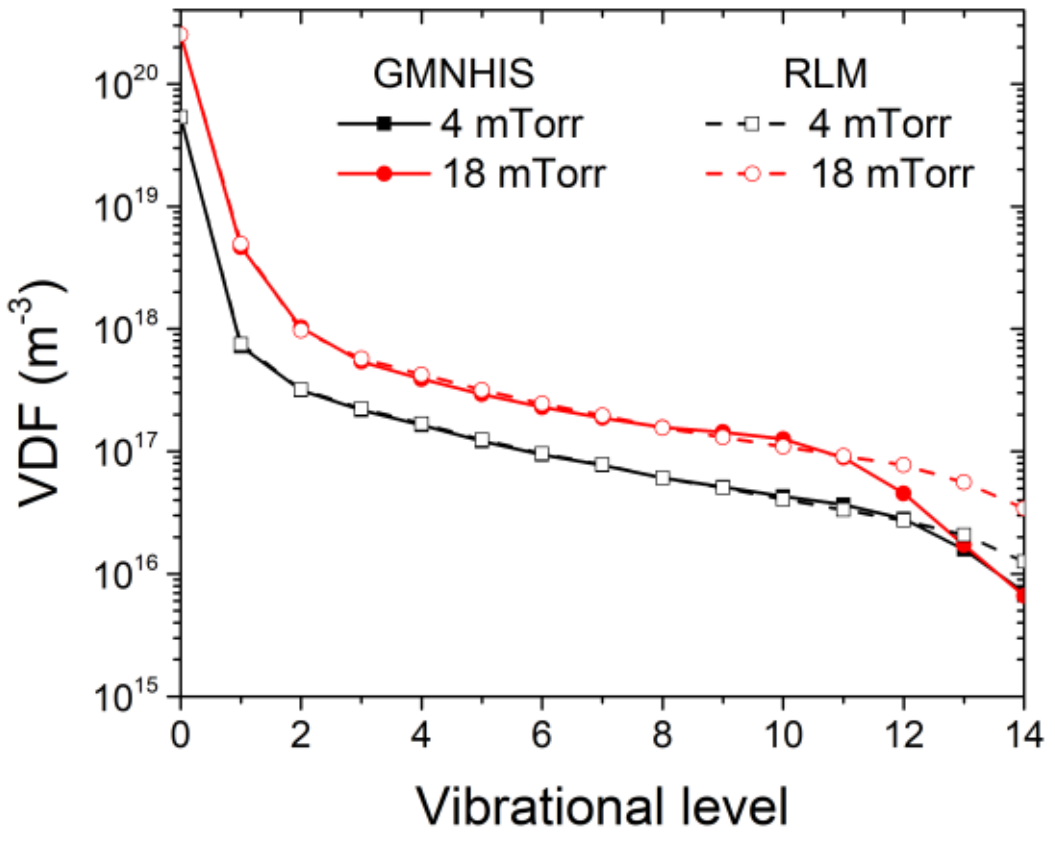

Figure 6. VDFs obtained using full GMNHIS and the RLM (reduced set without taking VT processes into account) for gas pressures 4 mTorr and 18 mTorr.

From Fig. 6 it is evident that VDFs obtained using full GMNHIS and the RLM are nearly identical except for the highest vibrational levels above 12. However, the highest vibrational levels do not contribute much into the negative ion production rate [12,16], and, therefore, it is not important to predict VDF for these levels accurately. As a result RLM, given by Eq. (5) can accurately reproduce the VDF except for the highest vibrational levels above 12.

In summary, the Vibrational Distribution Function (VDF) is found to be weakly dependent on the specific shape of the wall-repopulation probability function $\gamma_{\upsilon'\to\upsilon}$. This allowed substitution of actual $\gamma_{\upsilon'\to\upsilon}$ to a simple step function

$$\gamma_{\upsilon'\to\upsilon} = (1/\upsilon')\theta(\upsilon' - \upsilon) .$$

Having a simple function for $\gamma_{\upsilon'\to\upsilon}$ allows for analytical solution of a set of linear equations for the VDF produced by electron excitation and wall collisions. We apply Reduced Linear Model (RLM) for VDF of $H_2$ molecules in Negative Hydrogen Ion Sources (NHIS) at low pressures. The resulting RLM of vibrational kinetics for density of vibrationally excited levels $n_{H_2(\upsilon)}$ is

$$k_{\mathrm{WR},\upsilon} n_{\mathrm{H_2}(\upsilon)} = S_\upsilon + \frac{1}{\upsilon+1}\sum_{k=\upsilon+1}^{N} S_k .$$

Here the left-hand side describes the effective volumetric loss (in m$^{-3}$s$^{-1}$) of excited molecules at the wall surface, $k_{\mathrm{WR},\upsilon} = \left[\frac{\Lambda^2}{D} + \frac{2V}{Av_{th}}\left(\frac{1+R_\upsilon}{1-R_\upsilon}\right)\right]^{-1}$, and the right hand side describes the effective source, as a weighted sum of $S_\upsilon = (k_{\mathrm{eV},0\to\upsilon} + k_{\mathrm{EV},0\to\upsilon}) n_{\mathrm{H_2}(0)} n_e$.

The obtained analytical solution shows that the density of molecules in vibrational level, $\upsilon$, $n_{\mathrm{H_2}(\upsilon)}$ is determined by the source rate of electron excitation of ground-state hydrogen molecules to that level, and by a weighted sum over source spectrum over higher levels $\upsilon' > \upsilon$ with a weight of $\frac{1}{\upsilon+1}$. This convenient analytical solution can be used as a simple verification test for complex full numerical models.

We showed that the RLM reproduces well the VDF obtained using the full set of reaction in the GMNHIS code, except for the highest vibrational levels at relatively high pressure. The discrepancy for the highest vibrational levels is because the process of vibrational-translational relaxation in collisions with ground-state $\mathrm{H_2}(\upsilon = 0)$ molecules is not taken into account in the RLM. However, the highest vibrational levels do not contribute much into the negative ion production rate [12,16], and, therefore, it is not important to predict VDF for these levels accurately.

**Acknowledgments**

This work was supported by US Department of Energy, China Scholarship Council (CSC), National Magnetic Confinement Fusion Science Program, China (Grant No. 2015GB114000), National Key R&D Program of China (Grant No. 2017YFE0300106) and Initial Research Funds for Young Teachers of Donghua University.

**Appendix A: Derivation of the analytical solution**

In order to solve Eq. (3) of the main text, we introduce the following notation:

$$g_\upsilon = k_{\mathrm{WR},\upsilon} n_{\mathrm{H}_2(\upsilon)}, \; G_\upsilon = \upsilon \sum_{\upsilon'=\upsilon}^{N} \frac{g_{\upsilon'}}{\upsilon'}, \; \beta_\upsilon = \frac{k_{\mathrm{WR},\upsilon}}{k_{\mathrm{WR},\upsilon} + k_{\mathrm{out}}}, \text{ and } T_\upsilon = \beta_\upsilon S_\upsilon .$$

Under these definitions,

$$g_\upsilon = G_\upsilon - \frac{\upsilon}{\upsilon+1} G_{\upsilon+1}, \tag{A1}$$

and Eq. (3) becomes

$$T_\upsilon = G_\upsilon - \frac{\beta_\upsilon + \upsilon}{\upsilon+1} G_{\upsilon+1}. \tag{A2}$$

Eq. (A2) is a two-term recurrence and can be solved by a standard method for such equations. Introducing $\alpha_\upsilon = (\beta_\upsilon + \upsilon)/(\upsilon+1)$, we have $G_\upsilon = \alpha_\upsilon G_{\upsilon+1} + T_\upsilon$.

Note that without outflow, $\alpha_\upsilon \equiv 1$ and the above equation is already an arithmetic progression. Introducing new variables $x_\upsilon = G_\upsilon \Big/ \prod_{k=\upsilon}^{N} \alpha_k$, $Q_\upsilon = T_\upsilon \Big/ \prod_{k=\upsilon}^{N} \alpha_k$, we obtain $x_\upsilon = x_{\upsilon+1} + Q_\upsilon$.

The appropriate solution is $x_\upsilon = \sum_{k=\upsilon}^{N} Q_k$, or $G_\upsilon = \prod_{n=\upsilon}^{N} \alpha_n \sum_{k=\upsilon}^{N} Q_k$.

With the aid of Eq. (A1), we obtain

$$\begin{aligned}
k_{\mathrm{WR},\upsilon} n_{\mathrm{H}_2(\upsilon)} &= g_\upsilon = G_\upsilon - \frac{\upsilon}{\upsilon+1} G_{\upsilon+1} = \prod_{n=\upsilon}^{N} \alpha_n \sum_{k=\upsilon}^{N} Q_k - \frac{\upsilon}{\upsilon+1} \prod_{n=\upsilon+1}^{N} \alpha_n \sum_{k=\upsilon+1}^{N} Q_k \\
&= T_\upsilon + \frac{\alpha_\upsilon (1+\upsilon) - \upsilon}{1+\upsilon} \prod_{n=\upsilon+1}^{N} \alpha_n \sum_{k=\upsilon+1}^{N} \frac{T_k}{\prod_{m=k}^{N} \alpha_m} = \beta_\upsilon \left\{ S_\upsilon + \frac{1}{\upsilon+1} \prod_{n=\upsilon+1}^{N} \alpha_n \sum_{k=\upsilon+1}^{N} \frac{\beta_k S_k}{\prod_{m=k}^{N} \alpha_m} \right\} \\
&= \beta_\upsilon S_\upsilon + \frac{\beta_\upsilon}{\upsilon+1} \left\{ \beta_{\upsilon+1} S_{\upsilon+1} + \sum_{k=\upsilon+2}^{N} \beta_k S_k \prod_{n=\upsilon+1}^{k-1} \alpha_n \right\}.
\end{aligned} \tag{A3}$$

The second term for $\upsilon = N$, and the third term for $\upsilon = N-1$ and $\upsilon = N$ should be set to zero.

In the final form,

$$k_{\mathrm{WR},\upsilon} n_{\mathrm{H}_2(\upsilon)} = \beta_\upsilon S_\upsilon + \frac{\beta_\upsilon}{\upsilon+1} \left\{ \beta_{\upsilon+1} S_{\upsilon+1} + \sum_{k=\upsilon+2}^{N} \beta_k S_k \prod_{n=\upsilon+1}^{k-1} \alpha_n \right\},$$

$$\beta_\upsilon = \frac{k_{\mathrm{WR},\upsilon}}{k_{\mathrm{WR},\upsilon} + k_{\mathrm{out}}}, \; \alpha_\upsilon = (\beta_\upsilon + \upsilon)/(\upsilon+1).$$

If outflow is neglected, then $\beta_\upsilon = 1$, $\alpha_\upsilon = 1$ and the RLM solution becomes

$$k_{\mathrm{WR},\upsilon} n_{\mathrm{H}_2(\upsilon)} = S_\upsilon + \frac{1}{\upsilon+1} \sum_{k=\upsilon+1}^{N} S_k \ .$$

Note that to account for the so far neglected VT relaxation in collisions with ground-state molecules, one only needs to appropriately modify the value of $k_{\mathrm{out}} = k_{\mathrm{out}}(\upsilon)$.